\documentclass[showpacs,pre]{revtex4}
\usepackage{amsmath}
\usepackage{amssymb}
\usepackage{graphicx}
\usepackage{dcolumn}% Align table columns on decimal point
\usepackage{bm}% bold math
\newcommand{\Sec}[1]{Section~\ref{sec:#1}}
\newcommand{\Fig}[1]{Figure~\ref{fig:#1}}

\newcommand{\Eq}[1]{Eq.\ (\ref{eq:#1})}
\newcommand{\Eqs}[2]{Eqs~(\ref{eq:#1})~and~(\ref{eq:#2})}

\def\omegab{\mbox{\boldmath $\omega$}}
\def\calbb{\mbox{\boldmath $\mathcal{B}$}}
\def\calub{\mbox{\boldmath $\mathcal{U}$}}
\usepackage{latexsym}

\begin{document}

%\ead{email address}

\title{Geometric Results for Compressible Magnetohydrodynamics}
\author{Wayne Arter}
\affiliation{EURATOM/CCFE Fusion Association, Culham Science Centre, Abingdon, UK. OX14 3DB
}
\date{\today}

\begin{abstract}
Recently, compressible magnetohydrodynamics~(MHD) has been elegantly formulated
in terms of Lie derivatives. This paper exploits the geometrical
properties of the Lie bracket to give new insights into
the properties of compressible MHD behaviour, both with and without feedback
of the magnetic field on the flow.
These results are expected to be useful for the solution of 
MHD equations in both tokamak fusion experiments and space plasmas.
\end{abstract}


\maketitle

\section{Introduction}\label{sec:intro}
The recent work~\cite{Wa13a} showed how the equations of
ideal, compressible magnetohydrodynamics may be elegantly formulated
in terms of Lie derivatives, building on the work of Helmholtz, Walen
and Arnold. For example,
the equation of magnetic induction in a compressible flow may
be formulated in terms of
a Lie derivative of a vector by introducing the field~$\tilde{\bf B}$
defined as the the magnetic field~${\bf B}$ divided by the mass density,
\begin{equation}\label{eq:induc}
\frac{\partial\tilde{\bf B}}{\partial t}=\mathcal{L}_{\bf u}(\tilde{\bf B})
\end{equation}
where $\mathcal{L}_{\bf u}$ is the Lie derivative with respect to the
flow field~${\bf u}$, $\tilde{\bf B}={\bf B}/\rho$ and $\rho$ is mass density.
%Making the barotropic assumption that pressure~$p(\rho)$,
The dynamical, potential vorticity equation
may also be put into the Lie derivative form~\cite{Wa13a}
\begin{equation}\label{eq:fullpv}
\frac{\partial \tilde{\boldsymbol{\omega}}}{\partial t}=\mathcal{R}+{\mathcal{L}}_{\bf u}(\tilde{\boldsymbol{\omega}})-{\mathcal{L}}_{\tilde{\bf B}}(\tilde{\bf J})
\end{equation}
where the potential vorticity $\tilde{\boldsymbol{\omega}}=\nabla\times{\bf u}/\rho$
and the potential current $\tilde{\bf J}=\nabla\times{\bf B}/\rho$. The
term~$\mathcal{R}$ vanishes either upon making the barotropic assumption
that pressure~$p(\rho)$ or sometimes in the isentropic approximation.  The
system of equations is completed by the mass conservation relation
\begin{equation}\label{eq:mcons}
\frac{\partial\rho}{\partial t}+\nabla\cdot (\rho {\bf u})=0
\end{equation}

%Observe that all the vectors in the steady version of \Eq{fullpv} then satisfy $\nabla\cdot(\rho{\bf F})={\bf 0}$.

This work expands on and extends the results in ref~\cite{Wa13a}.
Much of it concerns further applications of the results in ref~\cite{Wa13a}
that rely on the peculiar properties of the Lie derivative and so 
are mostly geometrical in nature. After \Sec{math} devoted to the underlying
mathematics, there are two sections on applications. The first~\Sec{appl1}
discusses the relationship between coordinate bases and steady solutions of the ideal MHD
equations. The second section of applications~\Sec{appl2} uses
the coordinate-invariant property of the Lie derivative to investigate
how new solutions may be generated by use of coordinate mappings.
\Sec{addphys} considers how additional physical processes 
such as diffusion may be included in the MHD model and examines their effects
in analogous fashion.
\Sec{tdep} explores time-dependent solutions, and \Sec{summary} provides a brief
summary and discussion of the value of the more abstract mathematical approach
to MHD problems.

Although the main emphasis is on results for compressible MHD, there is
inevitably some overlap with other work on incompressible MHD. Most
of the previously published work on the subject is more directly constructive, concerned
typically with calculating  a steady equilibrium corresponding to a specified
pressure distribution. See the books~\cite{dhaeseleer,schindler} which
are biased towards applications in 
laboratory plasmas and space plasmas respectively. Of the more abstract,
geometrical and topological analysis of MHD, the
book by Arnold and Khesin~\cite{arnoldkhesin} cites a comprehensive selection
of works prior to its publication, although
the paper of Woolley~\cite{Wo90math} is a notable exception, cf.\ the work of \Sec{appl1}.
More recently Bogoyavlenskij~\cite{Bo01Infi,Bo02Symm} and Cheviakov~\cite{Ch05Cons}, see also the textbook~\cite[\S\,5.3.6]{blumancheviakovsanco},
have studied the capabilities of transformations to generate new equilibria
from `old', cf.\ the work of \Sec{appl2}.

\section{Mathematics}\label{sec:math}
\subsection{Coordinate Bases}\label{sec:coord}
A set of three vectors~${\bf e}_i,\,\,i=1,2,3$, forms a basis in 3-D provided the vectors
are linearly independent at each point. The vectors are said to form a
coordinate basis if each may be parameterised by~$x^i$ such that the
$x^i$ may be used as a set of coordinates. When this is not the case,
the~${\bf e}_i$ are referred to as a `frame'~\cite[\S\,4.5]{fecko}.
A set of coordinate vectors may be produced by
a mapping~${\bf{x}}(\bar{x}^1,\bar{x}^2,\bar{x}^3)$ to
the usual Cartesian coordinate system~${\bf x}=(x^1,x^2,x^3)=(x,y,z)$
from curvilinear coordinates~${\bf \bar{x}}=(\bar{x}^1,\bar{x}^2,\bar{x}^3)$, viz.
\begin{equation}\label{eq:coordv}
{\bf e}_i = \partial {\bf x} / \partial \bar{x}^i
\end{equation}

In the context of numerical grid generation, the~$\bar{x}^i$ are the
equivalent of the reference or computational coordinates, as for example they might be
the nodes of a regular cuboidal mesh, whereas the corresponding~${\bf{x}}$
might be arranged to sample the interior of an irregularly shaped cavity as uniformly as possible.
Normally such a mapping is required to be non-degenerate, except possibly
at a small number of isolated singular points such as exemplified by the
origin in polar coordinates. Hence, almost everywhere,
it constitutes a diffeomorphism, ie.\ the function ${\bf{x}}(\bar{x}^1,\bar{x}^2,\bar{x}^3)$
is differentiable as functions of its arguments, and invertible, ie.\ to
each~${\bf{x}}$ there corresponds a unique~${\bf \bar{x}}$.
The generation of such maps is a standard procedure in numerical grid
generation~\cite{gridgenhbook}.
Diffeomorphisms may also be conceived of as generated by
the flow-field of a smooth, non-vanishing vector~${\bf u}({\bf x},t)$.
%However, this view-point is more relevant to the time-dependent dynamics
%which will be discussed in the second paper~\cite{Wa13d} of this series.

Now in any reasonable 3-D coordinate system, as explained in the next \Sec{Lie},
there is the
remarkable result that the Lie derivative of a vector may be written
\begin{equation}\label{eq:lieder}
{\mathcal{L}}_{\bf v}({\bf w})^i= w^j\frac{\partial v^i}{\partial x^j}-v^j\frac{\partial w^i}{\partial x^j}
\end{equation}
where here and throughout, the Einstein summation convention will be used. Further,
superfixes will always indicate vector components, not exponents.
It is therefore helpful to introduce the Lie bracket notation
for the Lie derivative
\begin{equation}
\mathcal{L}_{\bf v}({\bf w})=[{\bf v},{\bf w}]
\end{equation}
so that the two vectors~${\bf v}$, ${\bf w}$  appear on an equal
footing. Adopting this notation~\cite[\S\,4.5]{fecko}, the condition that
the~${\bf e}_i$ form a coordinate basis may be expressed as
\begin{equation}\label{eq:coordb}
[{\bf e}_i,{\bf e}_j]={\bf 0},\;\;\; \forall i,\;j
\end{equation}

Conditionally, the converse also applies, ie.\ if \Eq{coordb} is satisfied
then the vectors~${\bf e}_i$ form a coordinate basis.
The conditions are that the~${\bf e}_i$ are everywhere non-vanishing and
linearly independent vectors and the result also requires the applicability
of the Poincare lemma~\cite[\S\,9.2]{fecko}. The lemma determines when an
irrotational vector field may be represented as the gradient of scalar.
The lemma requires the domain of interest (or manifold) to be topologically
simple, and notably excludes a toroidal geometry with the conventional
assignment of poloidal and toroidal angles to the range~$[0,2\pi]$.

An illustration from magnetic confinement
physics in a torus, involves the electric field of constant amplitude
applied in the toroidal direction which, according to  Faraday's Law,
becomes irrotational in the limit where
the magnetic flux change vanishes. Now if the torus is
imagined to be of very large major radius, so that it is effectively straight,
directed in the $z$-direction say,
then the electric potential $\Phi_E \propto z$. However the corresponding expression
in the torus would have to be $\Phi_E \propto \phi$, where $\phi$ is the angle about
the major radius,
hence if a multi-valued potential is to be avoided by confining $\phi$~to~$[0,2\pi]$, special treatment is required at
$\phi=0=2\pi$. Operationally, as in this example, the Poincare lemma can 
often be `forced to apply' by suitable use of boundary conditions, but it must be
remembered that there is an underlying topological constraint.

Here, the inapplicability of the Poincare lemma is unfortunate since it conflicts
with the requirement that the vectors
be non-zero everywhere in order to form part of a basis. In this context,
the Poincare-Hopf theorem is relevant. The theorem
relates the number of zeroes of a vector
field to the topology of the  compact manifold on which it is
defined (and gives the `hairy-ball' theorem in the case of spherical surfaces).
% it is obviously necessary that it be non-zero everywhere,
There follows that the only compact coordinate systems
are to be found in a toroidal geometry.  The only way therefore to 
produce a coordinate basis for a compact geometry is to let the angular coordinates
range freely over a torus.

\subsection{Lie and Other Derivatives}\label{sec:Lie}
Since the derivation of the coordinate invariance of the Lie derivative~\Eq{lieder}
helps understanding of the scope of the mapping approach, in particular what is
meant by `reasonable' in the previous \Sec{coord}, it will be given here.
First, suppose that an arbitrary vector~${\bf v}$ has Cartesian components $v^j$ 
and curvilinear components~$\bar{v}^j$. If ${\bf i}_j$ are the vectors of the Cartesian
orthonormal basis, often written~$\{{\bf \hat{x}},{\bf \hat{y}},{\bf \hat{z}}\}$, then since a vector
is the same regardless of the coordinate system employed
\begin{equation}\label{eq:vector}
{\bf v}=v^j {\bf i}_j=\bar{v}^j {\bf e}_j
\end{equation}
or, on taking Cartesian components
\begin{equation}\label{eq:vecdef}
v^j = \frac {\partial x^j} {\partial \bar{x}^k} \bar{v}^k 
\end{equation}
\Eq{vecdef} is often used to define a vector~\cite{lovelockrund}, viz.\ as a set of
quantities which transforms between coordinate systems following the above rule. Note that, unlike
some texts, ref~\cite{lovelockrund} has no constraint that the mapping be orthogonal, so that it does not
have to be a rigid-body rotation or translation. Since it is not obvious that \Eq{lieder}
defines a vector in a general curvilinear system, an important role of the following
derivation is to establish
that ${\mathcal{L}}_{\bf v}({\bf w})^i$ transforms as \Eq{vecdef}. 
Using \Eq{vecdef} to express  ${\bf v}$ and~${\bf w}$ in component form, \Eq{lieder} yields
\begin{equation}\label{eq:lieder2}
{\mathcal{L}}_{\bf v}({\bf w})^i=
\frac{\partial x^j}{\partial \bar{x}^k}\bar{w}^k\frac {\partial} {\partial x^j} \left(\frac{\partial x^i}{\partial \bar{x}^l} \bar{v}^l \right)-
\frac{\partial x^j}{\partial \bar{x}^k}\bar{v}^k\frac {\partial} {\partial x^j} \left(\frac{\partial x^i}{\partial \bar{x}^l} \bar{w}^l \right)
\end{equation}
which using 
\begin{equation}\label{eq:partials}
\frac{\partial }{\partial x^j}=\frac{\partial \bar{x}^m}{\partial x^j}\frac{\partial }{\partial \bar{x}^m}
\end{equation}
and expanding the vector derivatives, gives for the first term on the right-hand side
\begin{equation}\label{eq:lieder3}
\bar{w}^k \frac{\partial \bar{v}^l}{\partial \bar{x}^m}
\left(\frac{\partial x^j}{\partial \bar{x}^k}\frac{\partial x^i}{\partial \bar{x}^l} \frac{\partial \bar{x}^m}{\partial x^j}\right)
+\bar{w}^k \bar{v}^l
\left(\frac{\partial x^j}{\partial \bar{x}^k} \frac{\partial \bar{x}^m}{\partial x^j} \frac{\partial^2 x^i}{\partial \bar{x}^m \partial \bar{x}^l} \right)
\end{equation}
Now the rules of partial differentiation imply that 
\begin{equation}\label{eq:kronecker}
\delta^k_m = \frac{\partial x^j}{\partial \bar{x}^m} \frac{\partial \bar{x}^k}{\partial x^j}
\end{equation}
where the Kronecker delta symbol~$\delta^k_m=1$ if $k=m$ and is zero otherwise.
Hence \Eq{lieder3} simplifies dramatically, and noting that the second term on the right-hand side is the
same apart from interchange of the indices~$k$ and~$l$, there follows that
\begin{equation}\label{eq:lieder4}
{\mathcal{L}}_{\bf v}({\bf w})^i=
\bar{w}^k \frac{\partial \bar{v}^l}{\partial \bar{x}^k} \cdot \frac{\partial x^i}{\partial \bar{x}^l}
+\bar{w}^k \bar{v}^l \frac{\partial^2 x^i}{\partial \bar{x}^k \partial \bar{x}^l} 
-\bar{v}^k \frac{\partial \bar{w}^l}{\partial \bar{x}^k} \cdot \frac{\partial x^i}{\partial \bar{x}^l}
-\bar{v}^k \bar{w}^l \frac{\partial^2 x^i}{\partial \bar{x}^k \partial \bar{x}^l} 
\end{equation}
The terms in the second partial derivatives cancel, so that
\begin{equation}\label{eq:lieder5}
{\mathcal{L}}_{\bf v}({\bf w})^i=
\left(\bar{w}^k \frac{\partial \bar{v}^l}{\partial \bar{x}^k}
-\bar{v}^k \frac{\partial \bar{w}^l}{\partial \bar{x}^k} \right) \cdot \frac{\partial x^i}{\partial \bar{x}^l}
\end{equation}
which establishes both that the Lie derivative is a vector under coordinate transformation and
that it has a coordinate invariant expression.

To underline just how remarkable a result this is, consider the expression for the divergence of~${\bf v}$.
Differentiating \Eq{vecdef} with respect to~$x^i$, 
\begin{equation}\label{eq:divij1}
\frac{\partial v^j}{\partial x^i}=\frac {\partial\bar{v}^k} {\partial x^i} \frac{\partial x^j}{\partial \bar{x}^k} +
\frac {\partial} {\partial x^i} \left(\frac{\partial x^j}{\partial \bar{x}^k}\right)\bar{v}^l 
\end{equation}
which using \Eq{partials} gives
\begin{equation}\label{eq:divij2}
\frac{\partial v^j}{\partial x^i}=
\frac{\partial \bar{v}^k}{\partial \bar{x}^l} \frac{\partial x^j}{\partial \bar{x}^k}\frac{\partial \bar{x}^l}{\partial x^i}+
\frac{\partial \bar{x}^l}{\partial x^i} \frac{\partial^2 x^j}{\partial \bar{x}^k \partial \bar{x}^l} \bar{v}^k 
\end{equation}
Setting $i=j$ and summing (contracting indices $i$ and~$j$), and using \Eq{kronecker}, gives
\begin{equation}\label{eq:div}
\nabla\cdot{\bf v}= 
\frac{\partial v^j}{\partial x^j}=\frac {\partial\bar{v}^k} {\partial \bar{x}^k}+
\frac{\partial \bar{x}^l}{\partial x^j} \frac{\partial^2 x^j}{\partial \bar{x}^k \partial \bar{x}^l} \bar{v}^k 
\end{equation}
Introducing the Jacobian of the transformation between the two coordinate systems as~$\sqrt{g}$,
defined as the determinant
\begin{equation}\label{eq:jac}
\sqrt{g}= \frac{\partial(x^1,x^2,x^3)}
{\partial(\bar{x}^1,\bar{x}^2,\bar{x}^3)}
={\bf e}_1.{\bf e}_2\times{\bf e}_3
\end{equation}
it may be shown~\cite[\S\,4.1]{lovelockrund}, introducing cofactors and
using elementary calculus, that
\begin{equation}\label{eq:jacd}
\frac{\partial \sqrt{g}}{\partial \bar{x}^i}= \sqrt{g} 
\frac{\partial \bar{x}^l}{\partial x^j} \frac{\partial^2 x^j}{\partial \bar{x}^i \partial \bar{x}^l}
\end{equation}
Hence \Eq{div} may be written
\begin{equation}\label{eq:div2}
\frac{\partial v^j}{\partial x^j}=\frac {\partial\bar{v}^k} {\partial \bar{x}^k}+
\frac{1}{\sqrt{g}} \frac{\partial \sqrt{g}}{\partial \bar{x}^k} \bar{v}^k
\end{equation}
often rewritten as
\begin{equation}\label{eq:divg}
\frac{\partial v^j}{\partial x^j}=
\frac{1}{\sqrt{g}} \frac{\partial (\sqrt{g} \bar{v}^k) }{\partial \bar{x}^k}
\end{equation}
It is worth noting that the above formulae for Lie derivative and divergence actually
apply in any number of dimensions.

Other analysis establishes the formula for the curl
operator in curvilinear geometry.
\begin{equation}\label{eq:curlv}
\{\nabla \times {\bf v}\}^i=\frac{1}{\sqrt{g}} e^{ijk} \frac{\partial({g_{kl}\bar{v}^l})}{\partial \bar{x}^j}
\end{equation}
where the metric tensor~$g_{ij}$ is described in the next \Sec{mettensor} and $e^{ijk}=e_{ijk}$
is the alternating symbol, taking values $1$, $-1$ or~$0$,
depending whether $(ikl)$ is an even, odd or non-permutation of $(123)$.

The vanishing of the Lie bracket of two basis vectors is almost immediate from their definition,
for suppose that a vector function~$x^j(\bar{x}^i)$ is used to generate two vector fields
\begin{equation}\label{eq:vecf12}
{\bf u} = \partial {\bf x} / \partial \bar{x}^i,\;\;\;{\bf v} = \partial {\bf x} / \partial \bar{x}^j
\end{equation}
then by \Eq{kronecker}, ${\bf u}.\nabla=\partial/\partial \bar{x}^i$, ${\bf v}.\nabla=\partial/\partial \bar{x}^j$.
Upon substituting in \Eq{lieder5}, it becomes linear in second partial derivatives
and vanishes because the order in which partial derivatives are taken does not matter.

Lastly, the following useful results concerning the Lie bracket are noted, viz.
\begin{equation}\label{eq:liebrid1}
[ \mu {\bf u}, \lambda {\bf v}]= \mu [{\bf u}, \lambda {\bf v}]- {\bf u} (\lambda {\bf v}.\nabla \mu)
=\mu \lambda [{\bf u}, {\bf v}] +{\bf v} (\mu {\bf u}.\nabla \lambda)
-{\bf u} (\lambda {\bf v}.\nabla \mu)
\end{equation}
and in particular, setting ${\bf u}={\bf e}_i$, ${\bf v}={\bf e}_j$, then
\begin{equation}\label{eq:liebrid2}
[ \mu {\bf e}_i, \lambda {\bf e}_j]= 
%\mu \lambda [{\bf e}_i, {\bf e}_j] 
{\bf e}_j (\mu \lambda_{,i}) -{\bf e}_i (\lambda \mu_{,j})
\end{equation}
where suffix~$,i$ is used to denote differentiation with respect to~$\bar{x}^i$.
From \Eq{liebrid2}, it follows that if $\lambda$ is a function of~$\bar{x}^i$ only,
$\mu$ is a function of~$\bar{x}^j$ only, then the Lie bracket vanishes, ie.\ if each~${\bf e}_i$
is scaled by an arbitrary function~$\lambda_i$ of~$\bar{x}^i$ only, then the modified~${\bf e}_i$
also constitute a basis wherever the~$\lambda_i$ are non-zero.

\subsection{Metric Tensors}\label{sec:mettensor}
The metric tensor~$g_{ij}$ is introduced so the elementary distance~$ds$
measured in the coordinate system~${\bf \bar{x}}$ is
\begin{equation}\label{eq:mettensor}
ds^2=g_{ij}d\bar{x}^i d\bar{x}^j
\end{equation}
(using the Einstein summation convention)
when it follows that
\begin{equation}\label{eq:gequal}
g_{ij}= {\bf e}_i\cdot{\bf e}_j
\end{equation}
The quantity~$g$ is defined by $g=\det(g_{ij})$ and it may be shown, consistent with \Eq{jac}
that $g$ equals the square of the Jacobian of the transformation between the coordinate systems.
Elementary theory of determinants leads to the result that, in 3-D,
\begin{equation}\label{eq:2grad}
{\bf e}_i=\sqrt{g}(\nabla \bar{x}^j \times \nabla \bar{x}^k)
\end{equation}
where $(ijk)$ is a permutation of~$(123)$, whence
\begin{equation}\label{eq:diveg}
\nabla\cdot \left(\frac{{\bf e}_i}{\sqrt{g}}\right)= 0
\end{equation}

Specialising temporarily to 2-D coordinate systems, conformal mappings
between complex variables~$z=x+iy$ and $\bar{z}=\bar{x}+i\bar{y}$  may be defined
by $z= f(\bar{z})$ where $f(\bar{z})=u(\bar{z})+iv(\bar{z})$ is an analytic
complex function. (Note that overbar does \emph{not} denote complex conjugate.)
Elementary complex analysis then shows that 
\begin{equation}\label{eq:cpartials}
\partial{u}/\partial{\bar{y}}=\partial{v}/\partial{\bar{x}},\;\;\;
\partial{u}/\partial{\bar{x}}=-\partial{v}/\partial{\bar{y}},\;\;\;
\end{equation}
Introducing the shorthand $u_y=\partial{u}/\partial{\bar{y}}$ and obvious
variants, the metric tensor may be written
\begin{equation}\label{eq:cpg}
g_{ij}=\left(\begin{matrix}
u_x^2+v_x^2 & u_x.u_y+v_x.v_y\\
u_x.u_y+v_x.v_y & u_y^2+v_y^2\\
\end{matrix}\right)
=\left(\begin{matrix}
u_x^2+u_y^2 & 0\\
0 & u_y^2+u_x^2\\
\end{matrix}\right)
\end{equation}
and since from the result immediately after \Eq{gequal}, $g=(u_x^2+u_y^2)^2$, it follows that
for conformal mappings, $g_{ij}=\sqrt{g}\delta_{ij}$ where
$\delta_{ij}$ is the Kronecker delta.

No such reduction is possible in 3-D however. For suppose there
is an orthogonal mapping such that~$ds^2=h_{(i)}^2 \delta_{ij} d\bar{x}^i d\bar{x}^j$
(the $h_i$ are known as Lam\'{e} coefficients),
then $g=h_1^2 h_2^2 h_3^2$ and
\begin{equation}\label{eq:mat3}
G_{ij}=\frac{g_{ij}}{\sqrt{g}}=\left(\begin{matrix}
\frac{h_1}{h_2 h_3} & 0 & 0 \\
0 & \frac{h_2}{h_1 h_3} & 0 \\
0 & 0 & \frac{h_3}{h_1 h_2} \\
\end{matrix}\right)
\end{equation}
leading to the system of equations
\begin{equation}\label{eq:gsys}
h_1=h_2 h_3,\;\;\;
h_2=h_1 h_3,\;\;\;
h_3=h_1 h_2
\end{equation}
which may easily be shown to have no solutions except for $h_1^2=h_2^2=h_3^2=1$.

\subsection{Example of Cylindrical Polars}\label{sec:cylpolars}
Further to illustrate the mathematical machinery of coordinate transformations
just introduced, consider the mapping to cylindrical polar
coordinates~$(R,\theta,Z)$ given by
\begin{equation}\label{eq:cpdef}
x=R \cos \theta,\;\;
y=R \sin \theta,\;\;z=Z
\end{equation}
The basis vectors~${\bf e}_i$ follow by differentiation as
\begin{equation}\label{eq:cpbasis}
{\bf e}_1 =\frac{\partial {\bf x}}{\partial R}= (\cos \theta, \sin \theta,0),\;\;\
{\bf e}_2 =\frac{\partial {\bf x}}{\partial \theta}= (-R\sin \theta, R \cos \theta,0),\;\;\
{\bf e}_3 = \frac{\partial {\bf x}}{\partial Z}={\bf i}_3
\end{equation}
where the important point is that while the basis is indeed orthogonal, it is
not orthonormal.  Further, to work with the~${\bf e}_i$ as vectors
in Cartesian space, it is best to express them as functions of~$(x,y,z)$, viz.
\begin{equation}\label{eq:cpbcart}
{\bf e}_1 =\left(\frac{x}{r}, \frac{y}{r},0\right),\;\;\mbox{ where } r=\sqrt{x^2+y^2};\;\;\
{\bf e}_2 = (-y, x,0);\;\;\
{\bf e}_3 = (0,0,1)
\end{equation}

It can be seen by direct computation from \Eq{cpbcart}, using \Eq{gequal}, that
the metric tensor
\begin{equation}\label{eq:mats}
g_{ij}=\left(\begin{matrix}
1 & 0 & 0\\
0 & r^2 & 0\\
0 & 0 & 1\\
\end{matrix}\right)
\end{equation}
and hence $\sqrt{g}=r$. The expressions \Eq{cpbcart} may then be
used to verify for
example, that $\nabla.({\bf e}_2/\sqrt{g})=0$, because
\begin{equation}\label{eq:cpdiv}
\nabla.\left(\frac{{\bf e}_2}{\sqrt{g}}\right)=
\frac{\partial}{\partial x} \left(\frac{-y}{r}\right)+
\frac{\partial}{\partial y} \left(\frac{x}{r}\right)
\end{equation}
and it is easy to show by differentiating $r^2=x^2+y^2$ that
\begin{equation}\label{eq:cprdiff}
\frac{\partial r}{\partial x} =\frac{x}{r},\;\;
\frac{\partial r}{\partial y} =\frac{y}{r},\;\;
\end{equation}
so that the two terms on the right-hand side of \Eq{cpdiv} cancel.
The curls~$\nabla\times ({\bf e}_i/\sqrt{g})$ may be calculated in a similar
manner to the divergence. There is no simple general identity and the results have
no particular pattern, viz.
\begin{equation}\label{eq:cpcurl}
\nabla\times\left(\frac{{\bf e}_1}{\sqrt{g}}\right)=(0,0,0),\;\;
\nabla\times\left(\frac{{\bf e}_2}{\sqrt{g}}\right)=(0,0,\frac{1}{r}),\;\;
\nabla\times\left(\frac{{\bf e}_3}{\sqrt{g}}\right)=\left(-\frac{x}{r^3},\frac{y}{r^3},0\right)
\end{equation}

\section{Applications Exploiting the Basis Property}\label{sec:appl1}
%For the analysis in this section, it is convenient to consider also the mass flux vector ${\bf F}=\rho {\bf u}$, for then
%the mass conservation equation
%\begin{equation}\label{eq:mcons}
%\partial\rho/\partial t+\nabla\cdot (\rho {\bf u})=0
%\end{equation}
%means that~${\bf F}$ in steady-state must be solenoidal like the other fields~${\bf B}$,
%${\bf J}$ and~$\omegab$. Further, ${\bf u}$ can be conceived of as~$\tilde{\bf F}$.
\subsection{Steady Kinematics}\label{sec:kinematics}
Steady kinematics of the magnetic field is easily discussed in the present context,
for it requires the vanishing of the 3-D Lie bracket
$[{\bf u},\tilde{\bf B}]={\bf 0}$.
(This problem is referred to as kinematic MHD because~${\bf u}({\bf x},t)$ is an
arbitrarily specified field, ie.\ not necessarily dynamically consistent.)
%(In the language of differential geometry, $\tilde{\bf B}$ is said to be a symmetry field of~${\bf u}$.)
For steady flow, solutions where ${\bf B} \propto$ mass flux~${\bf F}=\rho {\bf u}$
are of course
well-known, but the geometrical results of the present paper
also indicate that if ${\bf u}$ and $\tilde{\bf B}$ are different members of a
coordinate basis, then $\tilde{\bf B}$ will also be a steady solution.
%For when $\rho=\mbox{const.}$, this means that a coordinate basis where
%two of the vectors satisfy $\nabla\cdot {\bf e}_i=$ gives
The divergence constraint is easily
met (for steady flow) %, because then both vectors~${\bf F}={\bf u}$ and~${\bf F}=\tilde{\bf B}$
%satisfy $\nabla\cdot \rho {\bf F}=0$.
because \Eq{2grad} implies  $\nabla\cdot ({\bf e}_i/\sqrt{g})=0$.
Hence, if ${\bf e}_1$ and ${\bf e}_2$ are two members of a coordinate basis,
%then ${\bf B}={\bf e}_2/\sqrt{g}$ is a steady solution of the induction equation in the flux ${\bf F}={\bf e}_1/\sqrt{g}$,
then $\tilde{\bf B}={\bf e}_2/(\rho \sqrt{g})$
is a steady solution of the induction equation in the steady flow
${\bf u}={\bf e}_1/(\rho \sqrt{g})$ with density~$\rho({\bf x})$, and 
and vice versa, provided
\begin{equation}\label{eq:rhog}
%{\bf e}_1 \cdot {\bf e}_2 \times {\bf e}_3=
\sqrt{g} \rho =\mbox{const.}
\end{equation}

The representation of fields~${\bf B}$ and~${\bf F}$
as basis vectors is equivalent to the use of the Clebsch
representation~\cite[\S\,2.5]{roberts}, \cite[\S\,5.2]{dhaeseleer}
for solenoidal 3-D vector fields.
This is in general only a local result, and there are significant
restrictions, notably in toroidal geometry, on its application
globally. The non-applicability of the Poincare lemma means that a
twisted magnetic field in a torus may not be identified in general
with a coordinate basis vector defined in terms of poloidal and toroidal angles,
unless their restriction to $[0,2\pi]$~is dropped, as discussed in \Sec{coord}.

Assuming the validity of the above representation, it is
not strictly necessary, here and in the next section for
${\bf e}_1$ and ${\bf e}_2$ to be linearly independent vectors,
since the requirement is only that they commute, ie.\ their Lie bracket vanishes.
Hence ${\bf e}_1 = {\bf e}_2$ is allowed here and later.
However, if they are linearly independent, then $\tilde{\bf B}$ and
${\bf u}$ form part of a coordinate basis.
This then implies that in steady-state there is 
everywhere a non-vanishing electric field with potential~$\Phi$, so that
\begin{equation}
{\bf u}\times{\bf B}= \nabla\Phi
\end{equation}

\subsection{Equilibria 1}\label{sec:steady1}
It is also possible to treat magnetic equilibria satisfying
$[\tilde{\bf J},\tilde{\bf B}]={\bf 0}$ in a similar vein to
the previous section.  The vanishing Lie bracket is equivalent to
the MHD equilibrium relation
\begin{equation}\label{eq:equil}
\nabla p = {\bf J}\times{\bf B}
\end{equation}
when the density is constant, which because of \Eq{rhog} implies
also that $\sqrt{g} =\mbox{const}$. \Eq{diveg} then implies
that the basis vectors are solenoidal.

There is a result from ref~\cite[\S\,II.1B]{arnoldkhesin},
to the effect that in a toroidal geometry $\nabla\times{\bf B}$
and ${\bf B}$ must represent a coordinate basis in each surface of constant
pressure, provided $\nabla p$ does not vanish.
However, supposing that ${\bf J}\propto{\bf e}_3$, there is
an additional constraint on the $\{{\bf e}_i\}$, namely that
\begin{equation}\label{eq:curlcon0}
{\bf e}_3=\nabla\times {\bf e}_2 
\end{equation}
Although the constraint~\Eq{curlcon0} above looks relatively simple,
substituting with \Eq{coordv} produces a complicated nonlinear
constraint on the mapping.
However, in the related problem
where ${\bf J} \propto {\bf e}_2$, taking the curl of ${\bf e}_2=\nabla\times {\bf e}_2$
leads to a type of vector Helmholtz equation. Ref~\cite{gridgenhbook} indicates that
equations of similar complexity may be successfully solved numerically
to generate coordinate mappings.
However, it is further necessary that $\sqrt{g}$ be constant, and this
is a demanding extra constraint which makes it hard to prove even that
mappings exist which also satisfy~\Eq{curlcon0}, although see ref~\cite[\S\,3.6]{hazeltinemeiss}.
This motivates looking at using the extra freedom allowed by a spatially
varying~$\sqrt{g}$.

\subsection{Equilibria 2}\label{sec:steady2}
Following the discussion of the previous \Sec{steady1}, it is natural to
seek equilibrium fields of the form
\begin{equation}\label{eq:bjeqm}
{\bf B}=\frac{b^2(\bar{x}^1,\bar{x}^3){\bf e}_2}{\sqrt{g}},\;\;\;
{\bf J}=\frac{j^3(\bar{x}^1,\bar{x}^2){\bf e}_3}{\sqrt{g}}
\end{equation}
where it will be seen that the forms chosen are the most general of `coordinate basis'
type that guarantee that the two fields are both solenoidal. (This representation
appears easier to work with than that suggested in ref~\cite[\S\,II.1A]{arnoldkhesin}.) To ensure
that \Eq{bjeqm} represents an equilibrium, requires from \Eq{liebrid2} that
\begin{equation}\label{eq:cindep}
\frac{\partial (b^2/\sqrt{g})}{\partial \bar{x}^3}=
\frac{\partial (j^3/\sqrt{g})}{\partial \bar{x}^2}=0
\end{equation}
These relations are satisfied if
\begin{equation}\label{eq:gindep}
\sqrt{g}=b^2 j^3 F_1(\bar{x}^1)
\end{equation}
where substituting directly in \Eq{equil}, $F_1(\bar{x}^1)=(\partial p/\partial \bar{x}^1)^{-1}$.
The remaining constraint is that ${\bf J}=\nabla\times{\bf B}$,
which in curvilinear coordinates becomes
\begin{equation}\label{eq:genjdef}
\frac{j^3 \delta_3^i}{\sqrt{g}}
=\frac{1}{\sqrt{g}}
e^{ijk} \frac{\partial}{\partial \bar{x}^j} \left(\frac{g_{k2}{b}^2}{\sqrt{g}}\right)
\end{equation}
or using \Eq{gindep}
\begin{equation}\label{eq:genj}
j^3 \delta_3^i =
e^{ijk} \frac{\partial}{\partial \bar{x}^j} \left(g_{k2}\frac{1}{j^3}
\frac{d p}{d \bar{x}^1}
\right)
\end{equation}

It follows that 
the metric tensor of the curvilinear coordinate system must satisfy \Eqs{genj}{gindep},
where the functions $b^2(\bar{x}^1,\bar{x}^3)$, $j^3(\bar{x}^1,\bar{x}^2)$
and $p'(\bar{x}^1)$ are arbitrary (prime denotes differentiation with respect to function
argument) except only that $\sqrt{g}$ and~$g_{22}$ cannot change sign. Since the system has four equations and the $g_{ij}$
represent six unknowns, it is plausible that solutions exist.
Indeed, it will be seen from \Sec{cylpolars},
\Eq{cpcurl}, that for $i=3$ and $k=2$
\begin{equation}\label{eq:curlcon}
{\bf e}_i/\sqrt{g}=\nabla\times ({\bf e}_k/\sqrt{g})
\end{equation}
so that the corresponding two cylindrical polar basis vectors represent an
equilibrium of the form~\Eq{bjeqm} with $b^2=j^3=1$. This solution might be used
to initialise calculations aimed at solving the equation pair
\begin{eqnarray}\label{eq:curlcon2}
j^3 {\bf e}_3&=&\nabla\times (b^2{\bf e}_2) + b^2{\bf e}_2 \times \nabla\ln(\sqrt{g})\nonumber\\
\sqrt{g}&=&\frac{b^2 j^3}{d p/d \bar{x}^1}
\end{eqnarray}
by substituting for the~${\bf e}_i$ using \Eq{coordv}.
As mentioned in \Sec{steady1}, this substitution
leads to a system that resembles those solved successfully as grid generation problems.
Another way to proceed %constructively
might be to introduce a pseudo-displacement current term, cf.\ \Eq{maxj} below.
In either context, \Eq{genj} is probably best regarded as a consistency check
to test coordinate mappings obtained by other means.

%The key mathematics is the expansion of the Lorentz Lie bracket using \Eq{liebrid1}
%\begin{equation}\label{eq:llexp}
%[\tilde{\bf J},\tilde{\bf B}]=\frac{1}{\rho^2}[{\bf J},{\bf B}]+
%-\frac{1}{\rho^3}\left(
%{\bf B} ({\bf J}.\nabla \rho) -{\bf J} ({\bf B}.\nabla \rho)
%\right)
%\end{equation}
%Thus it is necessary to ensure that the last two terms on the right-hand side
%vanish if $\rho$ is allowed to be variable as required to be by~\Eq{rhog}
%in order to compensate for $\sqrt{g}$ variations.
%
%Use of the Lie formalism to determine stability of equilibria
%will be discussed elsewhere~\cite{Wa13d}.

%The hope is that stability analysis will be easier in such systems.

Note, as will also emerge in the next \Sec{appl2},
that the Lie derivative formalism means nonlinear force balance is
trivially satisfied, whereas the linear relation~${\bf J}=\nabla\times {\bf B}$
causes difficulties. All other approaches to the problem of
calculating fully 3-D magnetic equilibria described in ref~\cite[\S\,8.1]{dhaeseleer}
work the other way around, ie.\ they substitute for~${\bf J}$ in
the nonlinear equation which they then solve typically by a 
variational technique.

\subsection{Example Equilibrium Converse}\label{sec:egeqm}
Now, consider the converse of the equilibrium problem examined in the previous
\Sec{steady1} and \Sec{steady2}, namely suppose that an equilibrium has been found
and a corresponding coordinate basis is required. Such a basis may be helpful
because the computation of Lie derivatives therein should generally be much easier.
Hence, consider the simplest non-trivial equilibrium relevant to magnetic confinement
physics~\cite[\S\,II.1B]{arnoldkhesin}, viz. a sheared magnetic field with 
\begin{equation}\label{eq:simpeqm}
{\bf B}=B(x){\bf \hat{z}},\;\;\; {\bf J}=-B'(x){\bf \hat{y}}
\end{equation}
(It is here noted that this is obviously related to the equilibrium described
in the preceding \Sec{steady2}, and further that coordinate transformations
which establish the correspondence will be described in \Sec{appl2}.)

There is a brute force approach to discovering the mapping which underlies
an equilibrium field, namely to assume ${\bf B}$ and~${\bf J}$ are
proportional to two different basis vectors, and seek a third via the vanishing Lie bracket
property of a basis. Alternatively, the mapping may be guessed, as here,  as
\begin{equation}\label{eq:simpeqmap}
x=\bar{x},\;\;\;y=-\frac{\bar{y}}{B(\bar{x})},
\;\;\;z=\frac{\bar{z}}{B'(\bar{x})},
\end{equation}
with inverse
\begin{equation}\label{eq:simpeqinvmap}
\bar{x}=x,\;\;\;\bar{y}=-B(x)y,
\;\;\;\bar{z}=B'(x) z
\end{equation}
when from \Eq{coordb}, the resulting basis vectors in Cartesian coordinates are
\begin{eqnarray}\label{eq:seqmbasis}
{\bf e}_1 &=& \frac{\partial(x,y,z)}{\partial \bar{x}} = \left(1, \frac{\bar{y}B'}{B^2}, -\frac{\bar{z}B''}{B'^2} \right)
=  \left(1, \frac{yB'}{B}, -\frac{zB''}{B'} \right) \\
{\bf e}_2 &=& \frac{\partial(x,y,z)}{\partial \bar{y}} = \left(0, -\frac{1}{B},0 \right) \\
{\bf e}_3 &=& \frac{\partial(x,y,z)}{\partial \bar{z}} = \left(0, 0, \frac{1}{B'} \right)
\end{eqnarray}
From \Eq{jac}, $\sqrt{g}=1/(BB')$, hence
\begin{eqnarray}\label{eq:seqmfld}
\frac{{\bf e}_1}{\sqrt{g}} &=&  \left(BB', -y B'^2, -zBB'' \right) \\
\frac{{\bf e}_2}{\sqrt{g}} &=&  \left(0, -B', 0 \right) = {\bf J}\\
\frac{{\bf e}_3}{\sqrt{g}} &=&  \left(0, 0, B \right) = {\bf B}
\end{eqnarray}
and it may be verified that all three~${\bf e}_i/\sqrt{g}$ are solenoidal.
It turns out that density which must satisfy $\rho\propto 1/\sqrt{g}$ so the above
formulae for ${\bf B}$ and~${\bf J}$ are consistent
with~$[\tilde{\bf B},\tilde{\bf J}]={\bf 0}$ is
given by $\rho\propto BB'$, hence is proportional to the equilibrium pressure gradient.

\subsection{Equilibrium with Flow}\label{sec:eqmflow}
There is the possibility that all three Lie brackets in the ideal MHD equations vanish.
For each field,
the divergence constraint may be easily met by scaling the basis by~$\sqrt{g}$.
Starting with the induction equation, suppose
\begin{equation}
{\bf B}={\bf e}_2/(\sqrt{g}),\;\;\; {\bf u}={\bf e}_1/(\rho \sqrt{g})
\end{equation}
then a steady-state with $\omegab \propto {\bf J} \propto {\bf e}_3$ is possible
provided \Eq{curlcon} is satisfied with $i=3$ and $k=2$,
not forgetting the requirement that $\rho \sqrt{g}=\mbox{const.}$ As in \Sec{steady1}
and \Sec{steady2}, 
${\bf e}_1$, ${\bf e}_2$ and ${\bf e}_3$ need not be linearly independent vectors,
and indeed two or more of them could be equal, but if they are independent, then
$\tilde{\bf B}$, $\tilde{\bf J}$ and ${\bf u}$ form a coordinate basis.

\section{Application to Complex Geometry}\label{sec:appl2}
\subsection{Invariance Properties}\label{sec:invar}
This section pursues the question as to what can be learnt about
MHD in complex geometries which are diffeomorphic to a Cartesian
geometry, from solutions in the Cartesian geometry. It provides a certain
amount of justification for the `rule of thumb' that such solutions
will be physically similar to those in the simpler geometry.

To focus ideas, suppose that the compressible magnetic induction
equations~\Eqs{induc}{mcons} have been solved
in Cartesian geometry and solutions for ${\bf B}$,
$\rho$  are available as functions of Cartesian coordinates~${\bf x}$
and possibly time~$t$ also. %in certain specific cases.
Consider first the expression from \Eq{divg} for the divergence of the
magnetic field vector in general geometry
\begin{equation}\label{eq:divb}
\nabla\cdot {\bf B} = \frac{1}{\sqrt{g}}\frac{\partial (\sqrt{g} \bar{B}^i)}{\partial \bar{x}^i}
\end{equation}
Since ${\bf B}$ must be solenoidal in the new coordinate system, it
is evident that if the components of~${\bf x}=(x,y,z)$ are to be identified
with~$(\bar{x}^1,\bar{x}^2,\bar{x}^3)$, then those of~${\bf B}=B^i=(B_x,B_y,B_z)$
must be identified with~$\sqrt{g} \bar{B}^i,\;\;i=1,2,3$.

%Similarly for the current
%${\bf J}$ which because of its definition as~$\nabla \times {\bf B}$ 
%also has to be divergence-free.
%$\tilde{\bf B}$ and $\tilde{\bf J}$.
%Now, for the Lie derivative in the Lorentz force term

Now, for the Lie derivative in the induction equation~\Eq{induc} term
to be identified in the way just
described requires similar separate identifications for
$\tilde{\bf B}$ and ${\bf u}$.
For this to be possible whilst
satisfying the divergence-free constraint on the field
requires $\rho \propto 1/\sqrt{g}$, cf.\ \Sec{appl1}. It will be seen
that this relation is, fortunately, consistent with mass conservation \Eq{mcons},
provided that $u^i$ is identified with~$\bar{u}^i$.
%Given this equivalence, the identification of the other two Lie derivative terms 
%in \Eqs{induc}{fullpv} follows.

To spell out the preceding, the mass conservation
equation may serve as an example for the others. In general
coordinates, \Eq{mcons} is
\begin{equation}\label{eq:massconsg}
\frac{\partial \bar{\rho} }{ \partial t} = -\frac{1}{\sqrt{g}} \frac{\partial (\sqrt{g} \bar{\rho} \bar{u}^i)}
{\partial \bar{x}^i}
\end{equation}
compared to the equation in Cartesians
\begin{equation}\label{eq:masscons}
\frac{\partial \rho }{ \partial t} = -\frac{\partial (\rho u^i)}{\partial x^i}
\end{equation}
%so that when the variable $\bar{\rho}=\rho \sqrt{g}$ is introduced, it becomes \Eq{masscons} becomes
%\begin{equation}\label{eq:massnog}
%\frac{\partial \rho }{ \partial t} =  \frac{\partial ( \bar{\rho} u^i)}
%{\partial \bar{x}^i}
%\end{equation}
%This last equation is of identical form to the mass conservation
%equation expressed in Cartesian coordinates.
%since it contains no metric information.
Hence when a solution~$\rho$ to \Eq{masscons} is found (as a function of Cartesian
coordinates), a solution in curvilinear coordinates~$\bar{\rho}$ follows in a flow
with contravariant velocity components~$\bar{u}^i$ set equal to
Cartesian components and position vectors $\bar{x}^i$ set
equal to Cartesian positions, namely that given
by~$\bar{\rho}=\rho/\sqrt{g}$.
Note that is \emph{not} simply the original solution expressed in the
new coordinates because this would see functions such as $\rho({\bf x})$
written
as functions of $\bar{x}^i$ through the mapping~${\bf x}(\bar{x}^i)$,
as well as the obvious difference of the $\sqrt{g}$ factor in the
case of the density.
%$\rho(\bar{x}^i,t)=\rho(\bar{x}^i,t)/\sqrt{g(\bar{x}^i)}$.

To complete the precise
identification for (`compressible') conformal invariance,
the following formulae are required
\begin{equation}\label{eq:ident}
\bar{x}^i=(\bar{x}^1,\bar{x}^2,\bar{x}^3) \leftrightarrow x^i=(x,y,z),\;\;
\bar{u}^i=(\bar{u}^1,\bar{u}^2,\bar{u}^3) \leftrightarrow u^i=(u_x,u_y,u_z),\;\;
\bar{\rho}(\bar{x}^i,t)\leftrightarrow  \rho(x^i,t)/\sqrt{g(x^i)}
\end{equation}
Writing $B^i=\bar{B^i}\sqrt{g}$, the Lie bracket
involves vectors exemplified by~$\tilde{B}^i=B^i/\rho$, which are
invariant $\tilde{B}^i \leftrightarrow  \bar{\tilde{B}}^i$ since
$\bar{B}^i/\bar{\rho}=B^i/\rho$. It is also possible to talk of
`incompressible' conformal invariance in steady-state, wherein
the density is not scaled, but~${\bf u}$ is, like all the other vector
fields, scaled by a factor~$\sqrt{g}$.

\subsection{Conformal Invariance}\label{sec:conform}
The above invariance properties of the compressible magnetic induction equations may
be useful, but it would be clearly be far more significant if something
similar could be found for the dynamic problem. At first sight, the Lie bracket
formulation of the Lorentz force term seems to make this possible.
Extending the argument concerning~${\bf B}$ of the previous section, since ${\bf J}$ must remain
divergence-free under transformation, $\tilde{J}^i \leftrightarrow  \bar{\tilde{J}}^i$
is invariant, so the Lorentz bracket is invariant.
The remaining difficulty is the need to satisfy
\begin{equation}\label{eq:curlb}
\bar{J}^i=\frac{1}{\sqrt{g}} e^{ijk} \frac{\partial({g_{kl}\bar{B}^l})}{\partial \bar{x}^j}
\end{equation}
After substituting for $\bar{B}^i$ and~$\bar{J}^i$ in
terms of~$B^i$ and~$J^i$ (to ensure solenoidal fields), it
appears that the simplest way to make the identification for
current, assuming that all three
components of~${\bf B}$ are non-vanishing, is for
\begin{equation}\label{eq:gid}
G_{ij}=\frac{g_{ij}}{\sqrt{g}} =g_0\delta_{ij}
\end{equation}
for some constant~$g_0$.

The \Eq{gid} implies from \Sec{mettensor} that the map between coordinate
systems must be a conformal mapping for solutions to be equivalent in the
sense described in the previous \Sec{invar}. The property is known as homothetic conformal
invariance, and is possessed by for example Maxwell's equations~\cite[\S\,16.4]{fecko}.
Although \Eq{gid} represents a strong constraint because it will usually restrict
maps to two-dimensions by demanding~$g_{33}=0$, the theory of conformal mappings
is well developed, and they may be used to generate an infinite number of solutions
from one in Cartesian geometry. In fact, conformal invariance is a group property,
so a first calculation in any conformally related coordinate system may be  used to
generate all the others.

\subsection{Restrictions on Conformal Invariance}\label{sec:restrict}
There are sadly a number of significant restrictions on the invariance property
of the preceding \Sec{conform}.
\subsubsection{Boundary Conditions}\label{sec:bcs}
Firstly, boundary conditions cannot be neglected in the
point mapping process. Obviously since all the 
fields acquire a factor~$\sqrt{g}$, boundary conditions will change quantitatively
in many cases. However, common conditions such as vanishing or
periodic field may still be satisfied if the mapping is chosen
appropriately near the boundary. For example, vanishing normal
component, or vanishing normal derivative may be preserved if the
new coordinates are arranged to be normal to the boundary.

\subsubsection{Failure with Potential Vorticity}\label{sec:potvort}
Secondly, although the advective Lie bracket in \Eq{fullpv} may be seen
to be invariant if ${\bf u}$ is identified as in the induction equation case,
the fact that a~$\sqrt{g}$ multiplier is not needed, means that the
definition of potential vorticity is \emph{not} conformally invariant.
Thus conformal invariance only applies to magnetic equilibria defined by
the vanishing of the Lorentz Lie bracket, although these could contain flow
provided the flow was either dynamically negligible or balanced by pressure
forces.

\subsubsection{The Subtle Restriction}\label{sec:subtle}
Even with the above two restrictions, there is a further and highly
important restriction which arises so subtly
it is best illustrated by example.
Consider the simplest M\"{o}bius mapping of the complex plane,
viz. $z=1/\bar{z}$, which implies that
\begin{equation}\label{eq:mobmap}
x +i y = \frac{\bar{x} - i \bar{y}}{\bar{x}^2+\bar{y}^2}
\end{equation}
The point of selecting this mapping is that straight lines are sent into circles,
so that the simple sheared-field equilibrium~\Eq{simpeqm} will then occupy a finite
region of the plane and so might be realised at least approximately in a laboratory.

It is convenient first to relabel coordinates in \Eq{simpeqm} so that
\begin{equation}\label{eq:simpeqm2}
{\bf B}=B(x){\bf \hat{y}},\;\;\; {\bf J}=B'(x){\bf \hat{z}}
\end{equation}
Turning to the M\"{o}bius mapping, consider the line in the $(x,y)$-plane given by~$x=1/(2 x_r)$,
then the real part of \Eq{mobmap} shows that in the mapped plane
\begin{equation}\label{eq:mobeqn}
\bar{x}^2-2x_r \bar{x} +\bar{y}^2=0,\;\;\mbox{ or }\;\; (\bar{x} - x_r)^2+  \bar{y}^2= x_r^2
\end{equation}
which is the equation of a circle centered at~$(x_r,0)$ with radius~$x_r$.
Hence the infinite region in $0<x<1/(2 x_r)$ is sent to the interior of this circle.

The basis vectors (restricted now to the 2-D~$(x,y)$-plane),
may be calculated as in earlier examples as
\begin{equation}\label{eq:mobbasis}
{\bf e}_1 =\frac{\partial (x,y)}{\partial \bar{x}}=
\left(\frac{\bar{y}^2-\bar{x}^2}{\bar{r}^4}, \frac{2 \bar{x} \bar{y}}{\bar{r}^4}\right),\;\;
{\bf e}_2 =\frac{\partial (x,y)}{\partial \bar{y}}=
\left(-\frac{2 \bar{x} \bar{y}}{\bar{r}^4}, \frac{\bar{y}^2-\bar{x}^2}{\bar{r}^4}\right)
\end{equation}
where $\bar{r}^2=\bar{x}^2+\bar{y}^2$. It follows that
\begin{equation}\label{eq:matmob}
g_{ij}=\left(\begin{matrix}
\bar{r}^{-4} & 0 \\
0 & \bar{r}^{-4} \\
\end{matrix}\right)
\end{equation}
and hence $\sqrt{g}=1/\bar{r}^4$.  
The identification of $B^i$ gives $\sqrt{g}\bar{B}^2(\bar{x})= B(x)$, so
\begin{equation}\label{eq:mobb}
\bar{B}^2(\bar{x},\bar{y})= \bar{r}^4 B(\bar{x}),\;\;\; \bar{B}^1(\bar{x},\bar{y})=\bar{B}^3(\bar{x},\bar{y})=0
\end{equation}
and similarly
\begin{equation}\label{eq:mobj}
\bar{J}^3(\bar{x},\bar{y})= \bar{r}^4 B'(\bar{x}),\;\;\; \bar{J}^1(\bar{x},\bar{y})=\bar{J}^2(\bar{x},\bar{y})=0
\end{equation}
The relation that $\bar{\bf J}=\nabla\times \bar{\bf B}$ may be
verified direct from the definition of the curl operator, as a
consequence of the fact that $g_{11}=g_{22}=\sqrt{g}$ in this example.

\emph{But} the Lorentz force may be computed directly in the curvilinear system
using the formula
\begin{equation}\label{eq:Lforc}
{F_L}_i= \sqrt{g} e_{ijk} J^j B^k
\end{equation}
and it will seen be immediately that this is not expressible as the gradient of
a scalar~$\partial p/\partial x^1$ unless $\sqrt{g}$ is a function of~$\bar{x}^1$
only, ie.\ nearly all
orthogonal mappings are disallowed. The further restriction results from the 
fact that ${\mathcal{L}}_{\tilde{\bf B}}(\tilde{\bf J})={\bf 0}$
does not, from \Eq{liebrid1}, enforce the equilibrium relation~${\mathcal{L}}_{{\bf B}}({\bf J})={\bf 0}$
unless ${\bf B}.\nabla \rho= {\bf J}.\nabla \rho=0$. It will be seen that this
further restriction is however at least consistent with the assumption that~$p(\rho)$.

It has to be concluded that this `compressible' conformal
invariance is little more general that the `incompressible' variety, wherein
the resulting further constraint is,
similar to that above, namely
${\bf B}.\nabla \sqrt{g}= {\bf J}.\nabla \sqrt{g}=0$. Plus, if further
${\bf u}.\nabla \sqrt{g}= \omega.\nabla \sqrt{g}=0$, the entire MHD system
including pressure is conformally `incompressibly' invariant in steady-state.

A further cautionary note (applicable to either variety of invariant)
is provided by the resistive term  which is often
added to the induction equation, given by
\begin{equation}\label{eq:resind}
{\bf R}=\frac{1}{\rho}\nabla \times \frac{{\bf J}}{\sigma_E}
=\frac{1}{\rho}\nabla \times \frac{1}{\sigma_E} \left(\nabla \times \frac{1}{\mu}{\bf B}\right)
\end{equation}
where $\sigma_E$ is the electrical conductivity and $\mu$~is
the plasma permeability~(normally $\mu=\mu_0$, the free-space value).
It seems on first inspection that this term is conformally invariant, but
unfortunately this fact is
not generally useful because when trying to complete the identification,
it is found that the repeated curls in \Eq{resind} normally
force two different diagonal components of~$g_{ij}$ to be zero.

The use of conformal mappings above is to be contrasted with Goedbloed's
conformal mapping approach~\cite[\S\,16.3.3]{GKP}, which is
used directly to generate equilibria. In the present work, it is necessary for
a solution to have been generated first, then it can be mapped.
In this respect the present work more closely resembles ref~\cite{Ma12Fini},
but the treatment herein does allow for equilibria with flow.

\section{Additional Physics}\label{sec:addphys}
This section investigates in more detail than was permitted in ref~\cite{Wa13a},
how additional physics terms may be introduced into the simplest ideal
system. Electrical resistance or equivalently magnetic diffusivity
was already discussed in the previous section. Repeated use of \Eq{curlv} in
\Eq{resind} gives the additional term in the required component form.

\subsection{Pressure Forcing}\label{sec:pterm}
It is particularly interesting to see, since the barotropic
assumption may well be overly restrictive, what
is required in order to include the forcing term
\begin{equation}\label{eq:force}
\mathcal{R}=\frac{1}{\rho}\nabla\times\left(\frac{\nabla p}{\rho}\right)
=\frac{1}{2}\left(\nabla\frac{1}{\rho^2}\times\nabla p\right)
\end{equation}
In Cartesian components, this may be written
\begin{equation}\label{eq:forcec}
\mathcal{R}_{x^i}
=\frac{1}{2}\left(\frac{\partial(\rho^{-2}, p, x^i)}
{\partial(x,y,z)}\right)
\end{equation}
This representation is convenient because Jacobians transform very simply
into  general geometry. Writing $\bar{\rho}$ and
$\bar{p}$ for $\rho$ and~$p$ evaluated as functions of~$\bf \bar{x}$, \Eq{forcec}
becomes
\begin{equation}\label{eq:forceg}
\mathcal{R}^i
=\frac{1}{2\sqrt{g}}\left(\frac{\partial(\bar{\rho}^{-2}, \bar{p}, \bar{x}^i)}
{\partial(\bar{x}^1,\bar{x}^2,\bar{x}^3)}\right)
\end{equation}
It should be evident that \Eq{forceg} is simply a convenient way of summarising the
three different 2-D Jacobians which have to be added to the corresponding
components of \Eq{fullpv}.

It may be further deduced from \Eq{forceg}, as might be expected,
that this term is not conformally invariant in the `compressible' sense. This
property ultimately results from the presence of the gradient operator
in the momentum equation.

\subsection{Anisotropy}\label{sec:aniso}
The point is that if code is written to treat general geometry
then it is easily extended to treat  
plasmas' having an anisotropic tensor permeability~$\mu_{ij}$.

Recall that for a general medium, Maxwell's equations give
\begin{equation}\label{eq:maxj}
{\bf J}=\nabla \times {\bf H}+ \frac{\partial {\bf D}}{\partial t},
\;\;\; \mbox{ where } \;\;\; {\bf B} = \mu {\bf H}
\end{equation}
Normally there is neglect of the displacement current term (containing~${\bf D}$)
in \Eq{maxj}.
Now from the general geometry expression for the curl, it is evident that
a $g_{ij}$ is required exactly where the reciprocal of~$\mu_{ij}$ should appear in \Eq{maxj}.
Further in the resistive case, see \Eq{resind}, a further $g_{ij}$~factor
is needed exactly where the reciprocal of the electrical conductivity
tensor~$\sigma_{Eij}$ is needed. Thus, since all tensors are symmetric and the products
and inverses of
symmetric tensors are symmetric, it is necessary only to allow for two
different `$g_{ij}$' to appear in the aforementioned places, to allow
for plasma anisotropy.

\subsubsection{Invariance Properties}\label{sec:invprop}
The result of the preceding paragraph also has an interesting
converse implication, namely that a solution obtained in Cartesian
geometry with anisotropic plasma properties satisfying
\begin{equation}
g_{ij} =\frac{1}{\mu_{ij}}=\frac{1}{\sigma_{Eij}}
\end{equation}
will furnish an equivalent solution
in general geometry, even in the resistive case.
This result holds provided that inertia is negligible
and subject to the possible redefinition of boundary  conditions
much as discussed in \Sec{bcs}.

In relation to other work on invariance properties,
contrast ref~\cite{Ch05Cons} where instead the pressure tensor
is taken to be anisotropic, and ref~\cite{Ma12Fini} where an extraneous
force term is introduced.
If inertia is significant, then  the situation
becomes that described in ref~\cite{Wa13a} where metric tensors have
to be explicitly introduced throughout.

\section{Time Dependent Behaviour}\label{sec:tdep}
\subsection{Elementary Field Kinematics}\label{sec:ekinematics}
Given that steady fields ${\bf u}_0$, $\rho_0$ and~$\tilde{\bf B}_0$ have been found
which satisfy the kinematic MHD equations, it is interesting to see whether
these may be used to construct time-dependent solutions. 

The simplest form to try is 
\begin{equation}\label{eq:simpsoln}
\tilde{\bf B}= \lambda(\bar{\bf x},t) \tilde{\bf B}_0
\end{equation}
Evidently
\begin{equation}\label{eq:inducs1}
\frac{\partial (\lambda \tilde{\bf B}_0)}{\partial t}=
\tilde{\bf B}_0 \frac{\partial \lambda}{\partial t}
\end{equation}
and from \Eq{liebrid1}
\begin{equation}\label{eq:inducs2}
[{\bf u}_0, \lambda \tilde{\bf B}_0] = \tilde{\bf B}_0 {\bf u}_0 .\nabla \lambda
\end{equation}
Hence
\begin{equation}\label{eq:inducs3}
\frac{\partial (\lambda \tilde{\bf B}_0)}{\partial t}-
[{\bf u}_0, \lambda \tilde{\bf B}_0] =
\tilde{\bf B}_0 \left(
\frac{\partial \lambda}{\partial t}
-{\bf u}_0 .\nabla \lambda
\right)
\end{equation}
so that \Eq{simpsoln} is indeed a solution of \Eq{induc} provided 
$\partial \lambda/\partial t =  {\bf u}_0. \nabla \lambda$.

However, ${\bf B}$ must remain solenoidal, implying $\tilde{\bf B}_0  .\nabla \lambda=0$
which is an awkward condition to satisfy in general. However, supposing that
$\tilde{\bf B}_0 ={\bf e}_2$, then the magnetic field, which will initially 
be  divergence-free provided $\rho \sqrt{g}=\mbox{const.}$,  remains divergence-free provided
only that $\lambda$ does not depend on~$\bar{x}^2$. An explicit solution may now
be realised by supposing that ${\bf u}_0={\bf e}_1$, for then 
\begin{equation}\label{eq:inducs4}
\lambda = F (\bar{x}^1+t, \bar{x}^3)
\end{equation}
where $F$ is an arbitrary function. It may be shown similarly that a consistent time dependent
density~$\rho= \xi(\bar{\bf x},t) \rho_0$
may be found provided that also $\xi= F_2(\bar{x}^1+t, \bar{x}^3)$.

\subsection{Time Dependent Kinematics}\label{sec:tkinematics}
This section illustrates the application of the Lie
derivative approach to the magnetic induction equation.
To treat the time dependent case, it is helpful to move to a 4-dimensional
space~$({\bf x},t)$ and introduce 4-vectors~$\tilde{\calbb}=(\tilde{\bf B},0)$,
$\calub=({\bf u},1)$, so that the evolution equation for~$\tilde{\bf B}$
may be written as
\begin{equation}\label{eq:induclie}
{\mathcal{L}}^{-}_{\calub}(\tilde{\calbb})=0,
%\mbox{four-vector flow }\;\;{\calub}=({\bf U},1)
\end{equation}
(The minus superscript denotes that the opposite sign convention to \Eq{lieder}
is being used in the definition of the Lie derivative.)
Introduce Greek indices~$\nu$, $\iota$, $\kappa$, $\lambda$ to index the 4-vectors, so
that for example~$\nu$ takes values~$1,2,3,4$ and $x^4=t$, then by definition
\begin{equation}\label{eq:inducexp}
{\mathcal{L}}^{-}_{\calub}(\calbb)=
\mathcal{U}^\lambda\frac{\partial \tilde{\mathcal{B}}^\nu}{\partial {x}^\lambda}-
\tilde{\mathcal{B}}^\lambda\frac{\partial \mathcal{U}^\nu}{\partial {x}^\lambda}
\end{equation}

Analogous to the 3-D result presented in \Sec{Lie}, substituting
\begin{equation}\label{eq:inducliesoln}
\tilde{\mathcal{B}}^\nu=\partial x^\nu / \partial \bar{x}^\iota,\;\;\;\;
\mathcal{U}^\nu=\partial x^\nu / \partial \bar{x}^\kappa
\end{equation}
in \Eq{inducexp} leads to a vanishing Lie bracket.
Taking $\kappa=4$ and $\bar{x}^4=t$ and letting~$\iota=j=1,2,3$,
application of the results in \Sec{Lie} shows that a solution of the induction
equation in the flow~${\bf \mathcal{U}}=(\partial{\bf x}/\partial t,1)$ is
\begin{equation}\label{eq:Btwsoln}
\tilde{\mathcal{B}}^\nu=\Sigma_{j=1}^{3} \tilde{b}^j\frac{\partial x^\nu}{\partial \bar{x}^j}
\end{equation}
provided $\partial \tilde{b}^j/\partial t=0$.
Clearly \Eq{Btwsoln} constitutes a completely general representation of a 3-vector,
hence \Eq{Btwsoln} is a completely general solution to a vanishing Lie bracket.

The kinematic MHD equations are not however solved until an expression has been
produced for the density, and this is easily provided if the
mapping~$x^\nu=({\bf x}(\bar{x}^1,\bar{x}^2,\bar{x}^3,t),t)$ is regarded as
being produced by a flow, meaning that it reduces to the identity at $t=0$
and thereafter changes smoothly with time. For such a map, the well-known
Lagrangian solution to the continuity equation gives
\begin{equation}\label{eq:rhoL}
\rho({\bf x},t) =\frac{1}{\sqrt{g}} \rho(\bar{\bf x},0)
\end{equation}
where $\sqrt{g}$ is the Jacobian of the map,
\begin{equation}
\sqrt{g}=
\frac{\partial(x^1,x^2,x^3,x^4)}
{\partial(\bar{x}^1,\bar{x}^2,\bar{x}^3,\bar{x}^4)}
=\frac{\partial(x^1,x^2,x^3)}
{\partial(\bar{x}^1,\bar{x}^2,\bar{x}^3)}
%=\frac{\partial(\bar{\bf x})}{\partial{\bf x})}
\end{equation}
the second equality following because $x^4=\bar{x}^4$. (Contrast \Eq{rhoL}
with \Eq{ident} where
because $\rho$ appears as a reciprocal in the vector fields, the $\sqrt{g}$~factor
is inverted relative to \Eq{rhoL}.)

Provided mass conservation is satisfied, an initially divergence free~${\bf B}$
satisfying~\Eq{inducexp} will remain solenoidal, hence the magnetic field solution of the induction
equation may be written
\begin{equation}\label{eq:Bsoln}
{\bf B}({\bf x},t)=\frac{1}{\sqrt{g}} \Sigma_{j=1}^{3} b^j(\bar{\bf x})\frac{\partial {\bf x}}{\partial \bar{x}^j}
\end{equation}
The above result re-expresses the well-known 
Lagrangian solutions to the ideal induction equation~\cite[\S\,2.3]{roberts}.

Consider lastly the flow described in $\bar{\bf x}$ coordinates by
\begin{equation}\label{eq:3flow}
\frac{d\bar{x}^1}{dt}=\frac{d\bar{x}^2}{dt}=0,\;\;\frac{d\bar{x}^3}{dt}=1
\end{equation}
Since
\begin{equation}\label{eq:3flow2}
{\bf e}_3= \frac{\partial {\bf x}}{\partial \bar{x}^3}=\frac{\partial{\bf x}}{\partial t}
\end{equation}
\Eq{3flow} corresponds
to a motion in physical space
in the direction of the ${\bf e}_3$ coordinate, cf.\ ref~\cite{Wa13b}.
%that time dependent solutions of the induction equation for a given flow
%may be generated by a continuously time-varying diffeomorphism which
%initially begins as the identity mapping~\cite[\S\,2.3]{roberts}.

\subsection{Quasi-2-D Dynamics}\label{sec:2dtdepdyn}
In ref~\cite{Wa13a}, a time dependent compressible MHD solution was
found given by $\tilde{B}^1=\tilde{B}^2=0,\;\;\tilde{B}^3=1$ (so that magnetic field 
and density evolution are tightly coupled),
subject to the restriction that all field quantities and the metric tensor depend only on
the other two coordinates~$\bar{x}^1,\bar{x}^2$.

\begin{figure}
\centerline{\rotatebox{0}{\includegraphics[width=10.0cm]{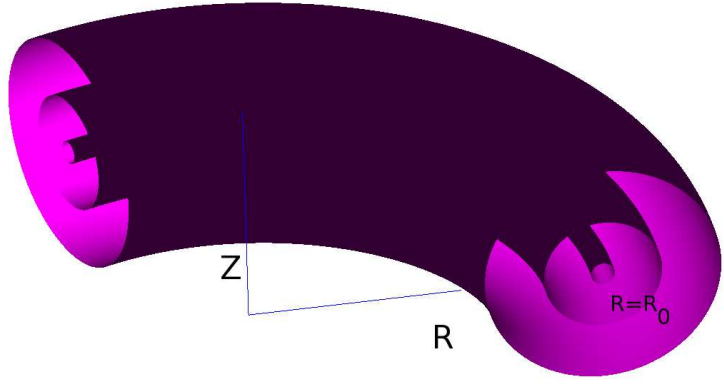}}}
\caption{Cutaway diagram of a torus, showing three different flux
surfaces~$\varrho(\psi,s,w)$ at constant~$\psi$, centred on $R=R_0$. 
(The diagram in fact shows the situation where  $\psi$ is independent of $s,w$.) As
explained in the text there is a correspondence between the coordinates~$\psi$,~$s$,~$w$
and $r$,~$\theta$,~$\phi$ respectively.
In a vertical plane, distance from the minor axis $R=R_0$ is usually denoted by~$r$
and the poloidal angle, usually denoted by~$\theta$, is measured around the minor axis.
The toroidal angle, usually denoted~$\phi$ is measured around the major axis~$R=0$.
\label{fig:torus2}}
\end{figure}

Further to explore the implications of this, introduce
generalised toroidal coordinates~$(\psi, s, w)$ (cf.\ $(r,\theta,\phi)$
as commonly employed in plasma physics~\cite{dhaeseleer}) so that 
\begin{equation}\label{eq:torc}
{\bf x}=(x,y,z)=\left(R_{+}\cos w, R_{+}\sin w, \varrho\sin s\right),
\end{equation}
where 
\begin{equation}\label{eq:rtorc}
R_{+}=R_0+\varrho(\psi,s,w) \cos s
\end{equation}
It will be seen from \Fig{torus2} that $\varrho(\psi,s,w)$ at constant~$\psi$
form  a set of nested
toroidal surfaces each having a major axis of~$R_0$. Introduce helical
coordinates~$(u,v)$ on each surface, so that $s=u-v/q(\psi)$,
$w=v+u/q(\psi)$, and write $\tilde{\varrho}(\psi,u,v)=\varrho(\psi,s,w)$.
Suppose that $\psi$ is rotationally symmetric about the $z$-axis
and satisfies the Grad-(Schl\"{u}ter)-Shafranov equation~\cite[\S\,8]{dhaeseleer},
ie.\ $\psi$~is a flux function for an equilibrium magnetic field, then
%provided $u$ and~$v$ are interpreted as generalised angles, meaning
the curves of \Eq{torc} as $v$ varies at constant $u$~and~$\psi$ are equivalent to lines
of the equilibrium field with helical pitch~$q(\psi)$. (Note that
$s$ and $w$, hence $u$ and $v$
need only be suitably periodic functions of the regular toroidal
angles~$\theta$ and~$\phi$. To define an equilibrium fully requires
defining these functions, but this is inessential for what follows.)
Since the metric tensor in a coordinate system $(\bar{x}^1,\bar{x}^2,\bar{x}^3)$
may be written
\begin{equation}\label{eq:gikd}
g_{ik}=\frac{\partial{\bf x}}{\partial \bar{x}^i}\cdot\frac{\partial{\bf x}}{\partial \bar{x}^k}
\end{equation}
taking $(\bar{x}^1,\bar{x}^2,\bar{x}^3)=(\psi, u,v)$ and using suffix~$i$ to denote
differentiation with respect to~$\bar{x}^i$, the components of~$g_{ik}$
are straightforwardly calculated as
\begin{equation}\label{eq:gpuv}
\left(
\begin{matrix} R_{+}^{2} w_\psi^{2}
         +\varrho^{2} s_\psi^{2}
         +(\tilde{\varrho}_\psi)^2 & R_{+}^{2} w_\psi w_u
         +\varrho ^{2} s_\psi s_u
         +\tilde{\varrho}_\psi \tilde{\varrho}_u & R_{+}^{2} w_\psi w_v
         +\varrho ^{2} s_\psi s_v
         +\tilde{\varrho}_\psi \tilde{\varrho}_v\\ 
         \cdot & R_{+}^{2} w_u^{2}
         +\varrho^{2} s_u^{2}
         +(\tilde{\varrho}_u)^2 & R_{+}^{2} w_u w_v
         +\varrho^{2} s_u s_v
         +\tilde{\varrho}_u \tilde{\varrho}_v\\ 
         \cdot & \cdot & R_{+}^{2} w_v^{2}
         +\varrho^2 s_v^{2}
         +(\tilde{\varrho}_v)^2\\ 
         \end{matrix}

\right)
\end{equation}
where the `dotted' entries may be deduced from the symmetry of~$g_{ik}$.
The usual rules for partial differentiation give
\begin{eqnarray}
\tilde{\varrho}_\psi &=& \varrho_\psi+\varrho_s s_\psi+\varrho_w w_\psi \nonumber  \\
\tilde{\varrho}_u &=& \varrho_s s_u+\varrho_w w_u  \label{eq:dsub}\\
\tilde{\varrho}_v &=& \varrho_s s_v+\varrho_w w_v \nonumber
\end{eqnarray}
The assumption of axisymmetric equilibrium is equivalent to $\varrho_w=0$, then
substituting \Eq{dsub} in \Eq{gpuv}, using the Reduce-algebra system~\cite{He05Redu}
gives
\begin{equation}\label{eq:gik}
\left(
\begin{matrix} R_{+}^{2} w_\psi^{2}
         +\varrho_\psi^{2}
         +2 \varrho_\psi \varrho_s s_\psi
         +\varrho_{+}^{2} s_\psi^{2} & \iota  R_{+}^{2} w_\psi
         +\varrho_\psi \varrho_s
         +\varrho_{+}^{2} s_\psi & 
         -\iota  \varrho_\psi \varrho_s
         -\iota  \varrho_{+}^{2} s_\psi
         +R_{+}^{2} w_\psi\\ 
         \cdot & \iota ^{2} R_{+}^{2}
         +\varrho_{+}^{2} & \iota  
         (R_{+}^{2}
           -\varrho_{+}^{2}
         )
         \\ 
         \cdot & \cdot & \iota ^{2} \varrho_{+}^{2}
         +R_{+}^{2}\\ 
\end{matrix}

\right)
\end{equation}
where $\varrho_{+}^2=\varrho^2+\varrho_s^2$ and $\iota=1/q$
has been introduced so that $s_{i}=(s_\psi,1,-\iota)$, $w_{i}=(w_\psi,\iota,1)$.

This $\bar{x}^k$~coordinate system has been chosen so that the equilibrium
field expected
in the tokamak confinement device may be expressed as $\tilde{B}^3=1$ ($\tilde{B}^1=\tilde{B}^2=0$),
but it will be seen that from the definition of $R_{+}$ in \Eq{rtorc}, the metric tensor does depend
on~$\bar{x}^3=v$ through $s=u-v\iota$.
By inspection, however, in the limit when $q$ is large and constant
or equivalently $\iota$ is negligible and constant (so $s_\psi=w_\psi=0$),
the metric tensor becomes
\begin{equation}\label{eq:gikc}
\left(
\begin{matrix} \varrho_\psi^{2} & \varrho_\psi \varrho_s & 
         -\iota  \varrho_\psi \varrho_s\\ 
         \cdot & \iota ^{2} R_{+}^{2}
         +\varrho_{+}^{2} & \iota  
         (R_{+}^{2}
           -\varrho_{+}^{2}
         )
         \\ 
         \cdot & \cdot & \iota ^{2} \varrho_{+}^{2}
         +R_{+}^{2}\\ 
         \end{matrix}

\right)
\end{equation}
thus $g_{ik}$ depends only on $\bar{x}^1$~and~$\bar{x}^2$. Hence a purely toroidal field,
ie.\ one tangent to circles about the major axis~$x=y=0$ of the torus,
allows for flux-compression solutions. Further, when $\varrho/R_0$ 
and $\varrho_s$ are both small,
the $s$-dependence of $g_{ik}$ is weak, so a helical field in a
circular torus with relatively large major radius is also in this category.
%$\phi \rightarrow  \theta - q \zeta$ shows that coordinates following
%a helical, shearless magnetic field with constant~$q$
%also have the necessary invariance property.
%%As shown by Schindler~\cite[\S\,5.2.3]{schindler}, $\partial g_{il}/\partial x^3=0$
%%also holds for helical coordinates.
The preceding limits illustrate two of the Killing vector solution
symmetries~\cite[\S\,5.2.4]{schindler}, whereas the third is simply invariance
in a Cartesian coordinate.

Lastly, if the coordinates~$s$ is redefined so that $s=u$, then the
metric tensor becomes
\begin{equation}\label{eq:giku}
\left(
\begin{matrix} (\varrho_\psi
           +\varrho_s s_\psi
         )
         ^{2}
         +\varrho ^{2} s_\psi^{2}
         +R_{+}^{2} w_\psi^{2} & \iota  R_{+}^{2} w_\psi
         +\varrho_\psi \varrho_s
         +
         \varrho_{+}^{2}
          s_\psi & R_{+}^{2} w_\psi\\ 
          \cdot & 
         \iota ^{2} R_{+}^{2}
         +  \varrho_{+}^2
         & \iota  R_{+}^{2}\\ 
        \cdot&\cdot& R_{+}^{2}\\ 
         \end{matrix}

\right)
\end{equation}
and $g_{ik}$ depends only on $\bar{x}^1$~and~$\bar{x}^2$ regardless
of the presence of magnetic shear. In this
coordinate system, however, lack of orthogonality makes it impossible in general to
arrange so that simultaneously $\tilde{B}^1=\tilde{B}^2=0$ and $\tilde{B}^3=1$.

\section{Summary}\label{sec:summary}
\Sec{math} has collated relevant results from differential geometry,
especially those concerning Lie derivatives, from a widespread 
mathematical literature. \Sec{appl1} and \Sec{appl2} then draw on these
results to derive solutions and solution methods for the ideal MHD equations
which are hopefully complementary to those presented in ref~\cite{dhaeseleer},
rather than simply `express old results in new language'.
The novelty of the Lie derivative approach is that coordinate basis
vector fields automatically cause the nonlinear terms in MHD to
disappear. The main bar to finding solutions is then the need to satisfy the
curl or `constitutive' relations ${\bf J}=\nabla\times{\bf B}$
and $\omegab=\nabla\times{\bf u}$.

In kinematic, compressible MHD, where
these constraints are not present, complete solution is possible
(see \Sec{tkinematics}), although this result was already known from
the Lagrangian approach. Suggestions are made for numerical solution
of the curl equations in \Sec{appl1} and \Sec{appl2}, but it seems that
for determining MHD equilibria,
the Grad-Shafranov equation %and its hydrodynamic equivalent still have
still has the advantage in most practical cases.

One possible practical improvement may result from the use
of coordinate bases, which simplify the calculation of Lie
derivatives, albeit in the example of \Sec{egeqm} at the expense
of introducing non-orthogonality. Linear stability analysis to investigate
this contention are outside the scope of the present work.

The reformulated equations at first sight appear to have significant
coordinate invariance properties, although \Sec{subtle} shows that
these are heavily qualified. The remaining parts of the present work
extend ref~\cite{Wa13a} by (i) considering additional physics effects
and how they might be modelled in the context of
differential geometry (\Sec{addphys}), and (ii) presenting
application of the flux-compression solution found in~\cite{Wa13a} (\Sec{2dtdepdyn}).

The current study has sought to apply results from 21st~Century
work on differential geometry to MHD. The mathematical notations and
concepts employed in the differential geometry literature have developed significantly
over the years. They may now seem to many excessively abstract, probably
because the subject has been strongly driven by Quantum Mechanics and
General Relativity applications, which require four or more
dimensions, and by other partial differential equation applications which require infinite
dimension. In 3-D, the coordinate-free approach can in fact be
unhelpful. The two curl equations exemplify this, for in the newer
language $\omegab=\nabla\times{\bf u}$ defines a 2-form in terms of a differential of a 1-form
or vector and is written $\omega=du$. However in ${\bf J}=\nabla\times{\bf B}$,
${\bf B}$ is already a 2-form, and so its dual has to be taken before
the `d' operator is applied, hence is written~$J=*d*B$.
Since this is often accompanied by a feeling of a
degree of uncertainty as to whether the operator on~$B$ should be thought
of acting in 3-D space or the 4-D Minkowski space-time where Maxwell's
equations are properly defined, the fact that there is no
indication above that the two curl equations have the same component form,
leads many to conclude that the more abstract approach just makes
a hard subject even harder.

It must however be pointed out that aspects of the modern, coordinate-free approach
are very valuable, for by having produced a more fundamental understanding of the
subject, it indicates in what circumstances analysis might lead to
physically useful results. The concept of the pull-back mapping
(even if it is misleadingly named, and although not used herein)
is also very useful for indicating
in the presence of coordinate mappings, precisely what functions of
which set of coordinates are being used. Without these latter positives,
the current paper would have been much weaker and more confused.

\section*{Acknowledgement}\label{sec:ackn}
I am grateful to Anthony J.~Webster for criticising the accessibility
of an early version of
the current work, and to John M.~Stuart for valuable advice on
the abstract mathematical background.
%This work, part-funded by the European Communities under the contract of
%Association between EURATOM and CCFE, was carried out within the framework
%of the European Fusion Development Agreement. The views and opinions
%expressed herein do not necessarily reflect those of the European
%Commission. This work was also part-funded by the RCUK Energy Programme
%under grant EP/I501045.
%This work was funded by the RCUK Energy Programme under grant EP/I501045 and
%the European Communities under the contract of Association between EURATOM and CCFE.
%The views and opinions expressed herein do not necessarily reflect those of
%the European Commission.
This work was funded by the RCUK Energy Programme grant number EP/I501045
and the European Communities under the contract of Association between
EURATOM and CCFE.
To obtain further information on the data and models underlying this
paper please contact PublicationsManager@ccfe.ac.uk.
The views and opinions expressed herein do not necessarily reflect
those of the European Commission.

%PRL These two lines to be removed
%\section*{References}
%\bibliographystyle{unsrt}
%PRL

%\bibliography{waynes,misc,new,warv,neuts}
\end{document}